\documentclass[letterpaper,amsmath,amssymb,aps,prl,reprint,superscriptaddress,footinbib,floatfix
]{revtex4-2}
\usepackage{geometry}
\usepackage[utf8]{inputenc}
\usepackage{lipsum}
\usepackage{graphicx}
\usepackage{dcolumn}
\usepackage{bm}
\usepackage{amssymb}
\usepackage{amsmath}
\usepackage{textcomp}
\usepackage{color}
\usepackage{mathtools}
\usepackage{units}
\usepackage{appendix}
\usepackage{microtype}

\begin{document}

\title{Expansion-Driven Self-Magnetization of High-Energy-Density Plasmas}

\author{K. V. Lezhnin}
\email{klezhnin@pppl.gov}
\affiliation{Princeton Plasma Physics Laboratory, 100 Stellarator Rd, Princeton, NJ 08540, USA}
\author{S. R. Totorica}
\affiliation{Department of Astrophysical Sciences, Princeton University, Princeton, NJ 08544, USA}
\affiliation{Department of Astro-fusion Plasma Physics (AFP), Headquarters for Co-Creation Strategy, National Institutes of Natural Sciences, Tokyo 105-0001, Japan}
\author{J. Griff-McMahon}
\affiliation{Department of Astrophysical Sciences, Princeton University, Princeton, NJ 08544, USA}
\author{M. V. Medvedev}
\affiliation{Department of Physics and Astronomy, University of Kansas, Lawrence, KS 66045}
\affiliation{Laboratory for Nuclear Science, Massachusetts Institute of Technology, Cambridge, MA 02139}
\author{H. Landsberger}
\affiliation{Department of Astrophysical Sciences, Princeton University, Princeton, NJ 08544, USA}
\author{A. Diallo}
\affiliation{Princeton Plasma Physics Laboratory, 100 Stellarator Rd, Princeton, NJ 08540, USA}
\author{W. Fox}
\affiliation{Princeton Plasma Physics Laboratory, 100 Stellarator Rd, Princeton, NJ 08540, USA}
\affiliation{Department of Astrophysical Sciences, Princeton University, Princeton, NJ 08544, USA}
\affiliation{Department of Physics, University of Maryland, College Park, MD 20742, USA}

\date{\today}

\begin{abstract}
Understanding plasma self-magnetization is one of the fundamental challenges in both laboratory and astrophysical plasmas. Self-magnetization can modify the plasma transport properties, altering the dynamical evolution of plasmas. Multiple high-energy-density (HED) experiments have observed the formation of ion-scale magnetic filaments of megagauss strength, though their origin remains debated. Here, we conduct 2D collisional particle-in-cell (PIC) simulations with a laser ray-tracing module for a fully self-consistent simulation of the plasma ablation, expansion, and magnetization. The simulations use a planar geometry, effectively suppressing the Biermann magnetic fields, to focus on anisotropy-driven instabilities. The laser intensity is varied between $10^{13}$ and $10^{14}$ W/$\rm cm^2$, which is relevant to HED and inertial fusion experiments where collisions must be considered. We find that above a critical intensity, the plasma rapidly self-magnetizes via an expansion-driven Weibel process, producing plasma beta of 100 ({$\beta = 8\pi k_B n_eT_e/B^2$}) and Hall parameter $\omega_{\rm ce}\tau_{e}>1$ within the first few hundred picoseconds. The magnetic field is sufficiently strong to modify plasma heat transport, and simulations with artificially suppressed magnetic field show noticeably different temperature profiles.

\end{abstract}

\maketitle

Understanding plasma magnetization is one of the fundamental challenges in both high-energy-density (HED)~\cite{StamperRipin1975,Nicolai2000} and astrophysical plasmas~\cite{KulsrudZweibel2008}. Magnetic fields are found in a wide variety of astrophysical systems, ranging from extremely strong $\sim 10^{11}$ T fields in compact astrophysical objects~\cite{Camenzind} to $\sim 10^{-9}$ T fields that permeate galactic and extragalactic regions~\cite{Widrow2002}, with the turbulent dynamo~\cite{Brandenburg2005}, Biermann battery~\cite{Biermann1950}, and Weibel instability~\cite{Weibel1959,Medvedev1999,Medvedev2006} serving as the primary theoretical hypotheses of plasma magnetization. The magnetization mechanisms are similar in HED plasmas produced in the laboratory. Biermann battery field generation is responsible for the plasma magnetization around the laser spot, {originating from plasma distributions with non-collinear electron temperature and density gradients (${\partial \bf{B}}/{\partial t} \propto \nabla T_e\times \nabla n_e)$}, producing azimuthal fields observed in laboratory astrophysics experiments on magnetic reconnection~\cite{Nilson2006,Zhong2010,Dong2012,Fiksel2014} and predicted in the inertial confinement fusion (ICF) simulations~\cite{Farmer2017,Sherlock2020}. Weibel magnetic fields {grow from noise via electromagnetic instability, drawing free energy from anisotropic particle velocity distributions. For a bi-Maxwellian distribution with temperatures $T_{\rm hot}>T_{\rm cold}$ and small anisotropy, they grow exponentially with a peak growth rate $\propto (T_{\rm hot}/T_{\rm cold}-1)^{3/2}$}. Weibel fields are commonly diagnosed in laser experiments involving counterstreaming plasmas~\cite{Fox2013,Huntington2015} and can drive collisionless shock formation in laboratory plasmas~\cite{Fiuza2020}.

\begin{figure}
    \centering
    \includegraphics[width=\linewidth]{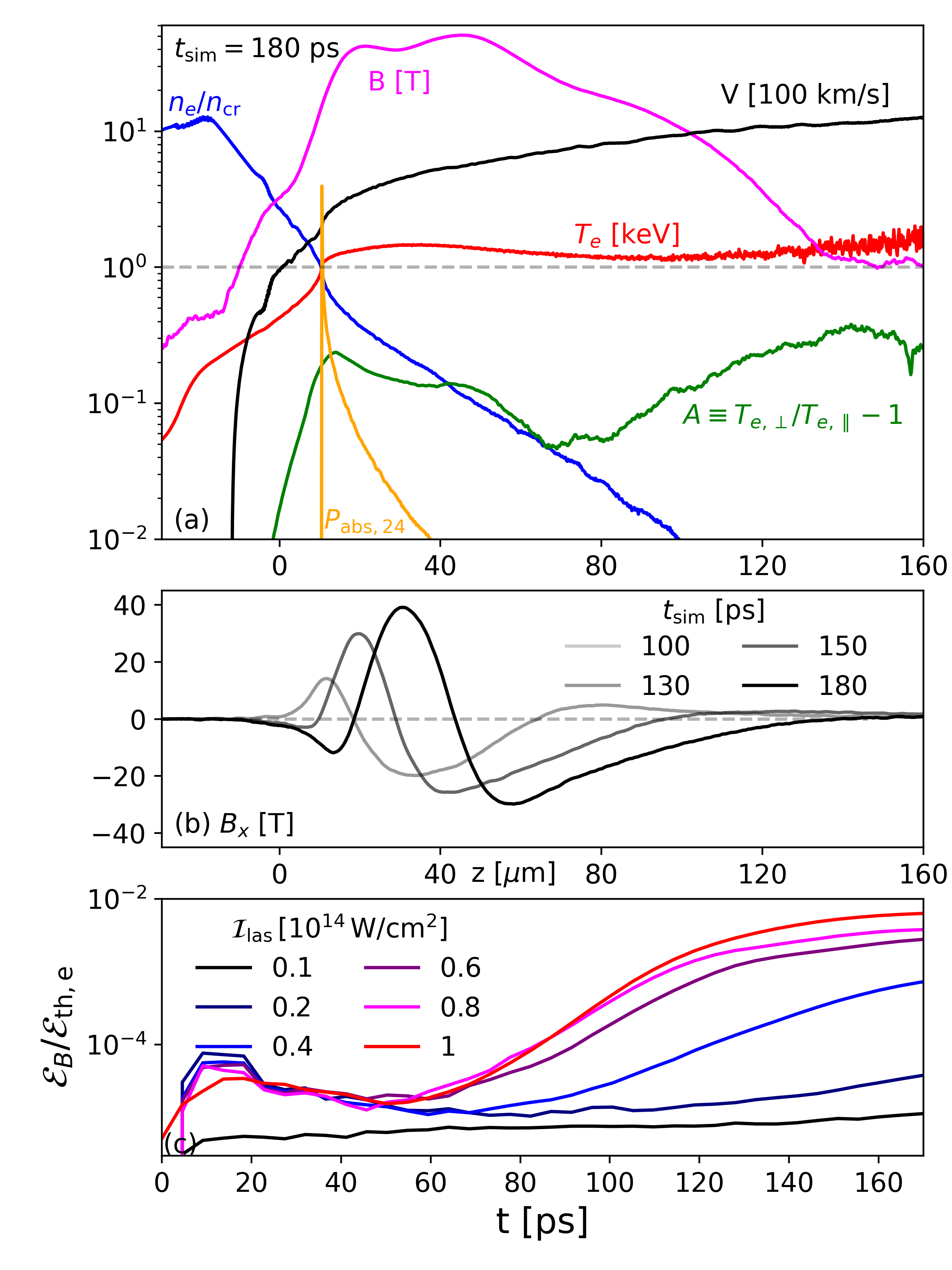}
    \caption{{(a) Plasma profiles and (b) $B_x$ field evolution in 1D  ablation by a laser with $I=10^{14}~\rm W/cm^2$.} Electron density (blue), temperature (red), expansion speed (black), magnetic field (magenta), laser absorption profile (yellow), {and electron temperature anisotropy $A\equiv T_{e,\perp}/T_{e,\parallel}-1$ ($\parallel$ is along z)} are depicted. (c) Evolution of magnetic to thermal energy ratio in the {$0\leq z \leq 250 \, \mu \rm m$} region for $I=10^{13} \rm W/cm^2$ (case considered in detail in Ref.~\cite{Lezhnin2024}) and $I=10^{14} \rm W/cm^2$ (primary case considered in this paper), along with intermediate values.}
    \label{fig:fig1}
\end{figure}

Despite a large body of experimental measurements and numerical models of plasma magnetization, the self-consistent magnetization of expanding plasma in HED laser experiments is still not fully understood. In one set of recent models~\cite{Schoeffler2014,Schoeffler2016,Schoeffler2020,Zhao2024}, kinetic simulations were initialized with specified profiles, either analytic or derived from radiation-hydrodynamics simulation, and showed that these conditions would produce a temperature-gradient-driven Weibel instability. Meanwhile, other works have shown that collisionless plasma expansion into vacuum could alternatively produce temperature anisotropy by preferential adiabatic cooling along the expansion direction, triggering {expansion-driven Weibel} magnetization ~\cite{Thaury2010a,Thaury2010b}. While both effects can drive Weibel instability (and, in principle, with identical growth rates for a given level of anisotropy), several features, including the location of the instability, the relevant plasma conditions, and the direction of anisotropy and the consequent polarization and structure of the Weibel magnetic field will differ. {The dominant mechanism of anisotropy development strongly depends on the expanding plasma profiles, requiring self-consistent modeling of laser-ablated plasma. The aforementioned studies, however, did not address the laser-driven ablation self-consistently, lacking either laser heating and self-consistent plasma profile generation \cite{Thaury2010a,Thaury2010b,Schoeffler2020,Zhao2024} or collisional transport effects relevant to HED plasmas \cite{Schoeffler2014,Schoeffler2016,Thaury2010a,Thaury2010b}.}

In this Letter, we present the first fully kinetic simulation of self-consistent laser-driven ablation, plasma expansion, and plasma magnetization relevant to the typical parameters of HED and ICF plasmas. Recently, in Refs.~\cite{Hyder2024,Totorica2024,Lezhnin2024,totorica2025particle}, we developed and validated a laser ray-tracing capability for PIC simulations within the code PSC~\cite{Germaschewski2016,Fox2018}, benchmarking it against the hydrodynamic codes FLASH~\cite{Fryxell2000,Tzeferacos2015} and RALEF~\cite{RALEF}, finding good agreement in the expanding plasma profiles in a strongly-collisional regime for relatively low intensities of $I =10^{10}-10^{13} \, \rm W/cm^2$. In the present work we build upon these results and extend to higher laser intensity and to 2D simulations, and show that the expanding plasma self-magnetizes above a critical laser intensity. We also show these self-generated fields modify the plasma expansion physics by restricting the heat transport along the expansion axis. The present simulations use a uniform transverse laser intensity profile {relevant to uniformly-driven HED systems such as direct drive capsules~\cite{craxton2015}}, and show the development of magnetic field consistent with an expansion-driven Weibel process, where the anisotropy is driven by preferential adiabatic expansion normal to the target surface, which is strong enough to overcome particle collisionality, sustaining sufficient anisotropy levels to drive the Weibel instability.

We begin by discussing laser-driven ablation in the case of a 1D simulation. Figure~\ref{fig:fig1} shows the result of a 1D PSC simulation, demonstrating magnetization of the laser-driven expanding plasma in the high-energy-density regime. The simulation setup is based on Ref.~\cite{Lezhnin2024} and is described in detail in the Appendix. In short, we consider an interaction of a laser with $I=10^{14}\, \rm W/cm^2$, $\lambda=1.064\, \rm \mu m$, coming from the right side of the box, with a fully ionized Al target of $n_e$=10 $n_{\rm cr}$ and {$30\,\rm \mu m$} thickness ($n_{\rm cr}=1.1\cdot10^{21}\, \rm cm^{-3}/\lambda_{\rm \mu m}^2$ is the critical density for the laser wavelength $\lambda_{\rm \mu m}=\lambda / 1 \, \rm \mu m$), and observe the expanding plasma evolution for 200 ps. Figure~\ref{fig:fig1}a presents plasma density, temperature, and expansion speed profiles, as well as laser heating (normalized to $10^{24}\,\rm erg/s/cm^3$), {electron temperature anisotropy}, and magnetic field profiles for $t=180$ ps. {For the purpose of this paper, we define the \textit{signed} anisotropy $A \equiv T_\perp / T_\parallel - 1$, where the parallel direction is along the target normal/laser/$z$ axis, and the perpendicular is in the plane of the target. The anisotropy can be positive or negative depending on what drives the evolution of the temperature components and will be discussed extensively below.} A magnetic field structure emerges just outside the critical surface (where $n_e(z) = n_{\rm cr}$), reaching {$50$} T and expanding towards the vacuum. {The electron temperature develops significant anisotropy $A>0$ at the same location, indicating the possible free energy source for magnetic field generation.} At the observed magnetic field strength, the electron gyrofrequency, $\omega_{\rm ce}$, is large enough, such that $\omega_{\rm ce} t_{\rm sim}>10$, indicating that the necessary condition for electron magnetization, $\omega_{\rm ce}t_{\rm sim}>1$, is satisfied. It should be noted that here we provide a conservative estimate of $\omega_{\rm ce}t_{\rm sim}$ from our simulation with ``heavy'' electrons due to reduced mass ratio, and one may expect $\omega_{\rm ce} t_{\rm sim}\gtrsim 10^2$ under realistic conditions. Figure~\ref{fig:fig1}b reveals the time evolution of the magnetic field structure. For the $B_x$ component, a decaying sine wave-like structure appears and grows around $z=${10-70 $\rm \mu m$}. Such a magnetic field structure is observed in $B_y$ as well ($B_z$ is suppressed in pure 1D simulations), and is qualitatively similar to $B_x$ and $B_y$ structures in 2D simulations, as we will show later. These oscillatory structures in magnetic fields are commonly associated with Weibel-like instabilities. Figure~\ref{fig:fig1}c tracks the time evolution of magnetic energy normalized to the plasma thermal energy in 1D simulations with various laser intensities. Here, we scanned over several values of the laser intensity, showing the development of self-magnetization with average magnetic energy fraction close to the $\sim 1\%$ level for laser intensities $I \ge 4\cdot10^{13} \, \rm W/cm^2$. 

\begin{figure}
    \centering
    \includegraphics[width=\linewidth]{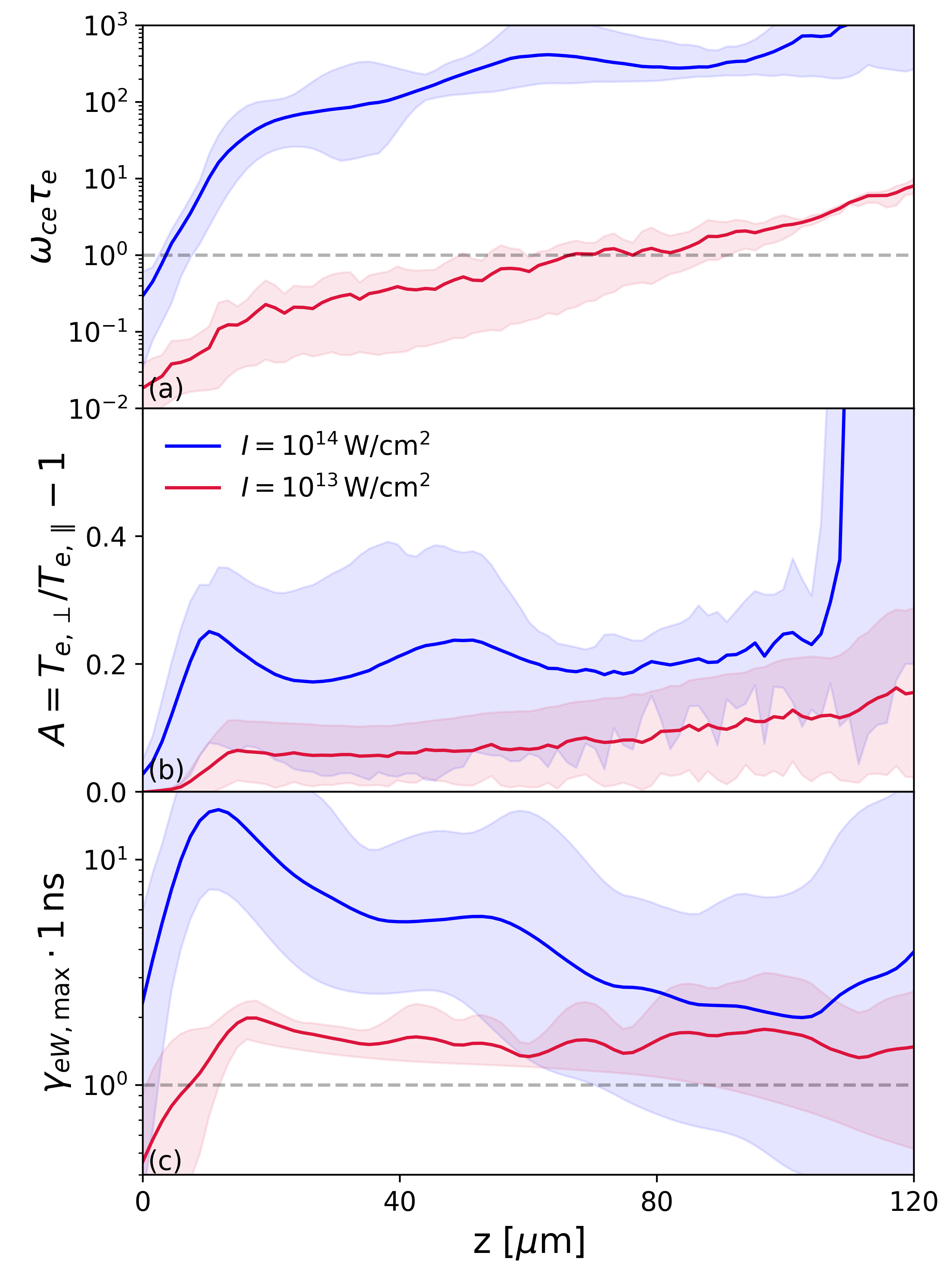}
    \caption{Transition from unmagnetized to magnetized plasma with laser intensity increase from $10^{13}$ to $10^{14}\rm \, W/cm^2$. (a) Hall parameter,  (b) electron temperature anisotropy, (c) collisionless electron Weibel growth rate for $I=10^{13}$ (red lines and shaded regions) and $10^{14}~\rm W/cm^2$ (blue lines and shaded regions) simulations {at t=100 ps}. The shaded regions denote the error bar given by the range of different numerical parameters (mass ratio, speed of light, dimensionality, particle resolution).}
    \label{fig:fig2}
\end{figure}

Figure~\ref{fig:fig2} shows how plasma self-magnetization compares between low and high laser intensities. It depicts profiles of (a) the Hall parameter, $\omega_{\rm ce}\tau_e$ ($\tau_e$ is the electron collisional timescale); (b) electron temperature anisotropy, $A$; and (c) normalized collisionless Weibel growth rate for two laser intensities: $I=10^{13}$ (red) and $10^{14}\, \rm W/cm^2$ (blue) {at t=100 ps}. The shaded regions summarize the results of a convergence study varying the reduced numerical parameters (reduced mass ratio, reduced speed of light, particles per cell, dimensionality). The collisionless electron Weibel growth rate calculated for the fastest-growing mode {in a bi-Maxwellian plasma with $A\ll 1$}, $\gamma_{eW0}$ (see, e.g., Ref.~\cite{davidson1972nonlinear}, Eq.~53), is given by:

\begin{equation}
\gamma_{eW0} = \frac{2\sqrt{2}}{3\sqrt{3\pi}}\frac{v_{th,e,\parallel}}{c}\omega_{\rm pe}\frac{A^{3/2}}{A+1},
\label{eq:eWeibel0}
\end{equation}

\noindent Here, $v_{th,e,\parallel}=\sqrt{k_BT_{e,\parallel}/m_e}$ is the electron thermal speed calculated for the parallel temperature, $\omega_{\rm pe}=\sqrt{4\pi e^2 n_e/m_e}$ is the electron plasma frequency, and $c$ is the speed of light in vacuum. {We note that this expression is valid for temperature-gradient-driven and expansion-driven electron Weibel instabilities in the linear stage for $A\ll 1$, see, e.g., Ref.~\cite{Schoeffler2020}, Eq.~2, and Ref.~\cite{Thaury2010b}, Eq.~20.} In Fig.~\ref{fig:fig2}a, the Hall parameter clearly transitions from below 1 to above 10 for an increased laser intensity, indicating a transition to plasma magnetization. The plasma magnetization is robust to varying the numerical parameters as demonstrated by the narrow shaded regions from the convergence study runs. In Fig.~\ref{fig:fig2}b, electron temperature anisotropy is present in both laser intensity cases, with higher levels of anisotropy in the $I=10^{14}~\rm W/cm^2$ case. We note that the anisotropy tends to decrease as the numerical parameters become more realistic (see Figure~\ref{fig:anisotropiesA} in the Appendix). Nevertheless, we find that all our simulations develop finite anisotropy with $A>0$, implying that the expansion direction is ``cold'' (i.e., $T_{e,\perp}>T_{e,\parallel}$). As is well-known from Weibel instability theory (see, e.g., Ref.~\cite{Kalman1968}), the fastest-growing modes align with the ``cold’’ direction, which in our case corresponds to modes with dominant $k_z$ and with transverse magnetic field polarization dominated by $B_x$ and $B_y$. In Fig.~\ref{fig:fig2}c, we see a distinct difference in the collisionless Weibel growth rates between the low and high intensity cases, with $\gamma_{eW0} \sim 10 \,\rm ns^{-1}$ in the high intensity case and $\gamma_{eW0} \sim 1 \,\rm ns^{-1}$ for the lower intensity. To summarize, all the features of Figure~\ref{fig:fig2} are consistent with an electron Weibel instability driving magnetic field generation in our simulations.

\begin{figure*}
    \centering
    \includegraphics[width=\linewidth]{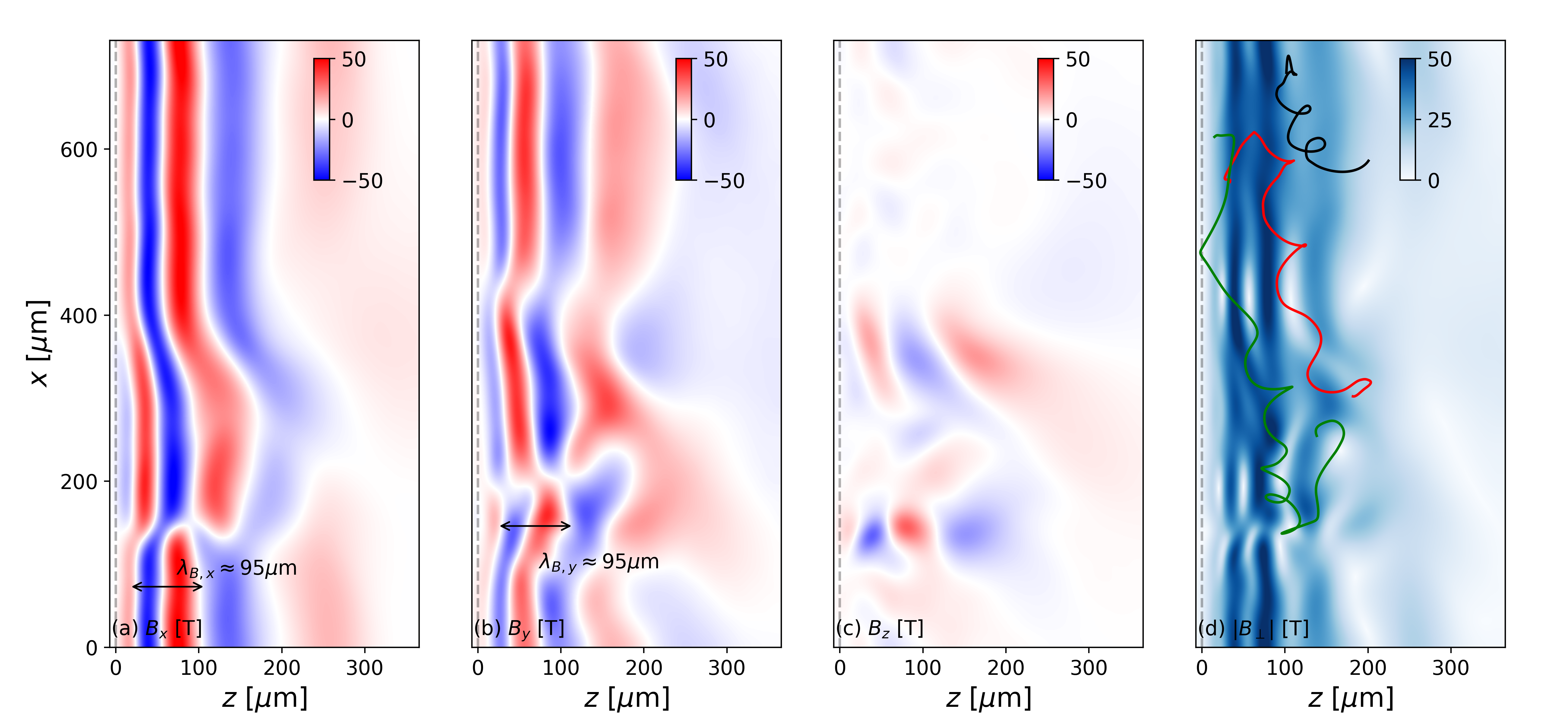}
    \caption{Magnetic field structure in expanding plasma at $t=300$ ps - (a)$B_x$, (b)$B_y$, (c)$B_z$, (d) $|B_\perp|$. Colored lines in (d) depict representative electron trajectories.}
    \label{fig:fig3}
\end{figure*}

Figure~\ref{fig:fig3} reveals the structure of magnetic field in a 2D simulation. This simulation is set up similarly to the 1D version but adds a transverse dimension along $x$, which has uniform laser intensity. 2D simulations open the possibility for all three components of the magnetic field to develop, which are shown in Figs.~\ref{fig:fig3}a-c. The magnetic fields reach $50$ T in magnitude at the end of the simulation ($t_{\rm sim}=300 \, \rm ps$), with the strongest components directed along the $x$ and $y$ axes. The typical wavelength of magnetic fluctuations along $z$ in our simulations revealed by the Fourier analysis is {$\lambda_{B,x}\approx \lambda_{B,y} \approx 95\, \mu \rm m$. This value is roughly consistent with the fastest-growing mode of Weibel instability~\cite{davidson1972nonlinear}, $\lambda_{\rm z, fast} = 2\pi\sqrt{3}c/\omega_{\rm pe}A^{-1/2}$, which may be estimated for our simulation conditions as $\lambda_{\rm z, fast}\sim 30~\mu \rm m$.} 2D simulations also yield inverse plasma beta of $\beta_e^{-1} \sim 10^{-2}$ and anisotropy levels approaching 0.1, in agreement with the 1D runs.

{The sign of the anisotropy that is developed and the resulting polarization of the magnetic filaments are the key evidence for the expansion-driven Weibel instability. In all our runs, including the extensive convergence tests (see Appendix), we observe that the coronal anisotropy $A$ is predominantly positive, which implies $T_{\perp}>T_\parallel$}; again, the ``cold'' direction is along the expansion axis $z$. We note that it contrasts the expectation for the temperature-gradient-driven Weibel instability, where thermal transport along $\nabla T_e$ (parallel to $z$) drives perturbations in $f_e(v_z)$, resulting in an increased parallel temperature, $T_{e,\parallel}$, and a negative value of the anisotropy parameter~\cite{Schoeffler2014,RamaniLaval1978}. On the other hand, in the expansion-driven Weibel case~\cite{Thaury2010a}, the expanding plasma adiabatically cools off along the expansion direction, thus decreasing the temperature along $z$. In our simulations, we observe $A>0$ across the whole simulation box, showing that the expansion-driven Weibel process dominates.

{The flow-driven development of electron temperature anisotropy may be demonstrated analytically. The second moment fluid equation for the electron temperature tensor in the absence of external forces, collisions, and heat fluxes may be written as $DT_{ij}/Dt + T_{ik}\partial u_j/\partial x_k+ T_{jk}\partial u_i/\partial x_k=0$, where $T_{ij}$ is the temperature tensor, $u_i$ is the flow vector, and $D/Dt$ is the convective derivative. This equation shows how general flow shears and compressions produce temperature anisotropy under collisionless conditions. Under the planar flow approximation ($u_z = az$, $a=\rm const >0$, $u_x=u_y=0$), $DT_{zz}/Dt=-2 a T_{zz},\, T_{xx}={\rm const}, \, T_{yy}=\rm const$. It is evident that the planar flow produces temperature anisotropy $A= (T_{xx}+T_{yy})/2T_{zz}-1>0$. (In the Appendix we justify the planar flow approximation and show how the result extends to radial flows from spherical capsules.) To further study analytically the competition of anisotropy development and collisional isotropization, we use a self-similar expansion model relevant to these flows~\cite{dorozhkina1998exact,bellei2012electron}, finding the anisotropy evolves as (see the Appendix for details):
\begin{equation}
    \frac{dA}{dt} =\frac{2C_S^2 t}{C_S^2t^2+L_0^2}(1-A)-\frac{6}{5}\left(1+\frac{1}{\sqrt{2}Z}\right)\nu_{ei}A.
    \label{eq:dadtfull}
\end{equation}
\noindent Here, $C_S$ and $L_0$ are the initial sound speed and plasma scale length. Solving this equation for typical HED plasma parameters (see Figure~\ref{fig:analytics} from the Appendix) yields a peak $A\approx 0.04$, which is sustained for several Weibel e-folding times during the simulation time of 200 ps. These findings are consistent with simulations, supporting the plausibility of the onset of expansion-driven electron Weibel instability in HED plasmas.}

\begin{figure}
    \centering
    \includegraphics[width=\linewidth]{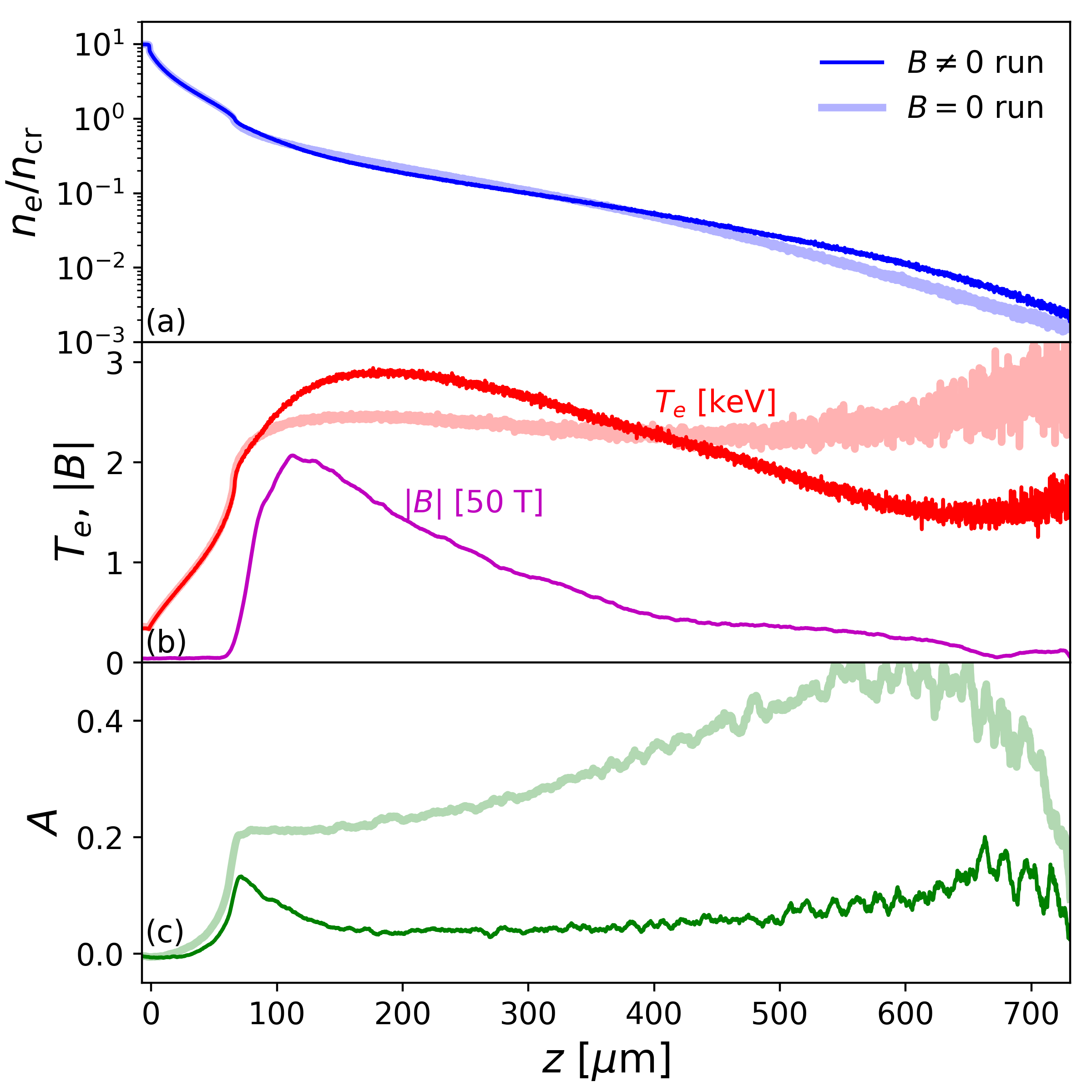}
    \caption{(a) Electron density, (b) electron temperature (in keV, red) and magnetic field (in 50s of Tesla, magenta), and (c) anisotropy profiles at $t=650$ ps comparing standard 1D simulations (solid lines) against 1D simulations with ``$B=0$'' (wide transparent lines). The dynamical effect of self-generated magnetic fields is evident in temperature and anisotropy profiles.}
    \label{fig:fig4}
\end{figure}

Figure~\ref{fig:fig4} shows how the self-generated magnetic field modifies the plasma expansion dynamics. Here, we ran two 1D simulations up to 650 ps, one with a regular particle pusher and another simulation with the $\mathbf{v \times B}$ Lorentz force inside the particle pusher turned off. We find that in the full simulation (solid lines), the temperature profile develops a spatial gradient in the target normal direction, whereas the ``$B=0$'' simulation (wide transparent lines) is highly isothermal (Fig.~\ref{fig:fig4}b). The magnetized simulation temperature rises about 20$\%$ above the unmagnetized case just outside the critical surface ({$z \approx 150~\mu \rm m$}), and falls to $\sim 20\%$ below the unmagnetized case further in the corona ({$z \approx 600~\mu \rm m$}). Such a peak is due to suppression of the outgoing heat flux by the magnetic field (Fig.~\ref{fig:fig4}b, magenta line). Inspecting the electron energy transport equation, we find that the heat flux contribution $|\nabla \cdot \mathbf{q}_e|$ is small compared to $|P_e\nabla \cdot \mathbf{v}|$ and $|\nabla \cdot (E_{e}\mathbf{v})|$ around the temperature peak ({$100~\mu \rm m \leq z \leq 300~\rm \mu m$}) in the magnetized case, but it is dominant in the ``$B=0$'' simulation for a similar laser heating power (difference in $Q_{\rm las}$ is within $30\%$). Here, $E_{e}$ and $P_e$ are the electron internal energy and pressure, respectively. Another notable difference between the magnetized and unmagnetized simulations is the electron temperature anisotropy, where the ``$B=0$''\,run develops a significantly larger level of anisotropy. This anisotropy is persistent throughout the simulation, as the magnetic fields are prohibited from amplifying via the instabilities. Magnetic fields, much like collisions, relax the anisotropy over time~\cite{Thaury2010a}, consistent with Figure~\ref{fig:fig4}c.

{Finally, an analysis of the magnetic structure and particle orbits also shows that the magnetic fields influence plasma heat transport. Perpendicular magnetic fields of sufficient strength (depicted in Figs.~\ref{fig:fig3}a,b,d) can trap electrons, preventing their expansion off the target, modifying the plasma heat transport. Figure~\ref{fig:fig3}d depicts three typical electron trajectories, which show gyrations around magnetic fields and preferential motion of electron gyrocenters perpendicular to the expansion direction. Moreover, we found that the gyroradii of 90\% of electrons in expanding underdense plasma are below the magnetic filament size (95 $\mu \rm m$). The analogous calculation for smaller laser intensity yields the opposite result, supporting the interpretation that the plasma becomes self-magnetized only above a threshold intensity.}

To qualitatively explain the emergence of Weibel magnetization in different laser-target interaction regimes, we propose a simple parameter based on the growth rate of a weakly collisional electron Weibel instability. Following Ref.~\cite{Schoeffler2020} and using the expression for the semi-collisional electron Weibel growth rate, we may calculate the ratio of the collisionless to collisional terms. The resulting ratio, after additional assumptions of fixed $\ln \Lambda=10$, fixed conservative anisotropy value of $A=0.01$, and plugging in the values for plasma temperature from the steady-state ablation model~\cite{Manheimer1982} (see also Ref.~\cite{Lezhnin2024}, Eq.~15) and estimating electron density by the critical density $n_{\rm cr}$, we can obtain a simple parameter that depends solely on laser intensity $I$, wavelength $\lambda$, and material properties (ion-to-proton mass ratio $\mu \equiv m_i/m_p$ and mean ionization $Z$):

\begin{equation}
    \Gamma \approx 2.8\, \frac{\mu^{2/3}}{Z^{5/3}}\left(\frac{\lambda}{1\,\mu \rm m} \right)^{11/3}\left(\frac{I}{10^{13}\rm \, W/cm^2} \right)^{4/3}.
    \label{eq:gamma_weibel}
\end{equation}

\noindent If $\Gamma$ is significantly larger than one, we expect development of electron Weibel instability. For a limited set of simulations with different materials (Al, C), laser wavelengths (0.532 and 1.064 $\mu \rm m$), and laser intensities ($10^{13}$-$10^{14} \, \rm W/cm^2$), Eq.~\ref{eq:gamma_weibel} correctly predicts the simulation outcome. We therefore can estimate the emergence of the Weibel instability based on a simple argument formulated with the $\Gamma$ parameter for the experimental regime of interest. Notably, $\Gamma \propto \lambda^{11/3}$, which suggests relative suppression of Weibel instability for 3$\omega$ beams ($\lambda=351$ nm) typically used in the ICF experiments at OMEGA and NIF. Still, for laser intensities $I\geq 8\cdot 10^{14}\, \rm W/cm^2$, $\Gamma \geq 1$ for Au and C, predicting an onset of Weibel process in ICF conditions.

It is instructive to compare our results with other studies on expanding plasma magnetization. Refs.~\cite{Schoeffler2020,Zhao2024} concluded that the parameter controlling the transition from Weibel to Biermann or from Weibel to no-Weibel is the Knudsen number, $\lambda_{e,\rm mfp}/L_T$. We calculate this parameter to be $\sim 0.3$ in our primary simulations, which was found sufficient to develop Weibel in Ref.~\cite{Zhao2024}. The difference in polarization of the self-generated magnetic fields between our study and Refs.~\cite{Schoeffler2020,Zhao2024} is due to the different geometry and plasma profiles used in their simulations.

Laser heating plays an important though indirect role in Weibel magnetogenesis. We conducted auxiliary simulations showing that the main role of laser heating is sustaining a high temperature in the expanding plasma to support fast plasma expansion and mitigate collisional isotropization. We tested that our implementation of the laser heating operator~\cite{Hyder2024} does not produce anisotropy by itself (see Fig.\ref{fig:anisotropiesB} in the Appendix), although we note that laser-driven anisotropy development is plausible in some scenarios~\cite{Bendib1998}.

In the present work, we considered a uniform transverse laser intensity profile, consistent with the conditions desired for direct-drive inertial fusion, which drives spherical targets with nearly uniform laser intensity ($|\delta I/I|\leq1\%$)~\cite{craxton2015}. While a detailed study of the effect of the laser spot profile is beyond the scope of the manuscript, we have also conducted preliminary studies with isolated laser spots. We find that the on-axis plasma magnetization still follows the expansion-driven Weibel scenario shown here, even as the gradient regions exhibit more complex magnetization behavior due to additional Biermann battery effect and oblique Weibel modes. {Nevertheless, the present results are significant for their application to uniform drive scenarios, as well as to directly illustrate the requirements for anisotropy generation in HED plasma conditions and to discriminate between commonly proposed anisotropy mechanisms~\cite{Bendib1998,Zhao2024,Thaury2010a,Davies2025}}.

In conclusion, this paper presents the first fully kinetic PIC simulation of laser-plasma interaction, expansion, and magnetogenesis under long-pulse laser conditions relevant to high-energy-density plasma experiments. We show that the plasma self-magnetizes above a critical intensity threshold due to expansion-driven Weibel instability, producing Hall parameters $\omega_{\rm ce} \tau_e > 1$ with concomitant modifications to plasma dynamics. {We support our findings with the analytical model of anisotropy generation by arbitrary flows in weakly collisional plasmas.} This investigation was made possible by improving our simulation capabilities associated with collisions~\cite{totorica2025particle}, laser ray-tracing ~\cite{Hyder2024}, and reinforced by extensive benchmarking ~\cite{Totorica2024,Lezhnin2024}. More broadly, we showed that fully kinetic simulations are feasible for high-energy-density plasmas, and may be of use for experimental design, interpretation, and cross-code benchmarking.

This work was supported by the U.S. Department of Energy under contract number DE-AC02-09CH11466. This work was supported by the Laboratory Directed Research and Development (LDRD) Program of Princeton Plasma Physics Laboratory. This research was supported in part by grant NSF PHY-2309135 to the Kavli Institute for Theoretical Physics (KITP). JGM acknowledges NSF support under Grant No. 2039656. MM acknowledges NSF support via grant PHY-2409249. The simulations presented in this article were performed on computational resources managed and supported by Princeton Research Computing at Princeton University. The authors thank J.R. Davies for fruitful discussions. 


\bibliographystyle{apsrev4-1}
\bibliography{magnetization}

\appendix

\section{Simulation setup}\label{sec:simsetup}

In our 1D simulations, we consider a 1D $292 ~\mu \rm m$-long simulation box, which is equal to $40 d_i$, where $d_i = c/\sqrt{4\pi n_{\rm cr} e^2/m_p}$ is the proton inertial length evaluated at the critical density. A solid target of fully ionized Al with an electron density of $10n_{\rm cr}$ is located between $z=-29.2~\mu \rm m$ and $0$. Here, $n_{\rm cr}=m_e\omega_0^2/4\pi e^2 \approx 1.1\cdot 10^{21} \left(\lambda [\rm \mu m]\right)^{-2}\, \rm cm^{-3}$ is the critical density for the laser wavelength $\lambda$. The laser is launched from the right side of the box, with a laser intensity $I=10^{14}\,\rm W/cm^2$ and a laser wavelength $\lambda = 1.064\, \rm \mu m$. The numerical grid is 2000 cells long, with $10^5$ particles per cell per species at a density of $n_e=n_{\rm cr}$. Common to PIC simulations, we use a reduced ion-electron mass ratio and speed of light; see Ref.~\cite{Fox2018} for details. A reduced proton-to-electron mass ratio $m_p/m_{e\ast}=$100 is used in the main set of simulations, as well as a reduced electron rest mass $m_{e\ast}c^2_\ast = 60 \, \rm keV$. See the Appendix of Ref.~\cite{Lezhnin2024} and the next section for additional information on the convergence studies. The simulation is conducted for $t_{\rm sim}=200$ ps.

The 2D simulations use a box size of 730 $\mu \rm m$ longitudinally by 292-1460 $\mu \rm m$ in the transverse direction for various runs, resolved by 50 cells per $d_i$ ($\approx$7 cells per micron) in each direction. A solid target of fully ionized Al with an electron density of $2.5n_{\rm cr}$ is located between $z=-29.2~\mu \rm m$ and $0$. $10^3$ electrons per cell at the critical density are used, with a convergence study performed with $10^4$ electrons per cell at the critical density. Periodic boundary conditions for fields and particles are applied in both directions. The simulation runs for 300 ps.

\begin{figure}
    \centering
    \includegraphics[width=\linewidth]{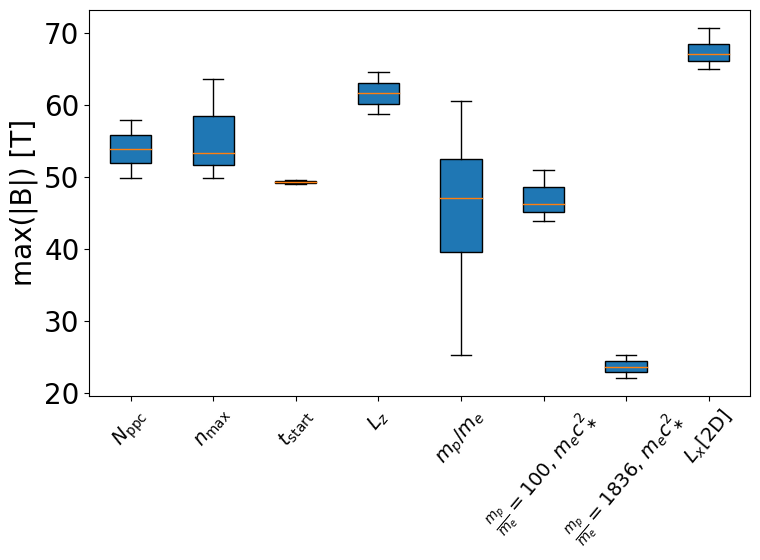}
    \caption{{\normalsize{Boxplot with a summary of numerical convergence tests. Number of particles per cell, peak density of the target, initial snapshot, longitudinal box size, ion-to-electron mass ratio, artificial speed of light, and transverse box size were varied. The resulting magnetic field is robust to variations in most reduced numerical parameters.}}}
    \label{fig:bscan}
\end{figure}

{
\section{Robustness study}\label{sec:robust}
In order to verify the validity of the assumptions used in our simulations, we conduct a series of auxiliary simulations that check the dependence of our results on numerical parameters. Below, we describe all the parameters checked. For the cases of specific interest for this manuscript, we provide additional figures; for the rest of the parameters, we list the results and refer the reader to the Appendix of Ref.~\cite{Lezhnin2024}.

Let us start by describing all the numerical parameters varied in 1D simulations. We check the dependence of our results on (1) box size, varying it from 292 to 1460 $\mu \rm m$, keeping the resolution fixed in terms of the number of grid cells per micron; (2) solid target density, varying it from 2.5 $n_{\rm cr}$ to 10 and 20 $n_{\rm cr}$, while keeping particle and grid resolutions fixed; (3) number of electrons per cell at the critical density, checking values of $10^4,\,10^5$; (4) initial plasma profiles from which the simulations were initiated — we use the FLASH code to produce electron density, temperature, and expansion speed profiles for the laser and target parameters of interest, taking snapshots at t=0.02 and 0.1 ns into the simulation and loading them into PSC; (5) proton-to-electron mass ratio parameter, which is varied from 100 to 1836; (6) electron rest energy (equivalent to the reduced speed of light), varied among 20, 60, and 512 keV. For 2D simulations, we additionally test the dependence of the simulation results on the transverse box size, varying it from $1d_e$ to $5,~20,~40,~80,~100,~120,~200 d_i$ (0.73 to 1460 $\mu \rm m$). All these simulations are included in Figure 2 of the main manuscript and depicted as the shaded regions. In general, all these simulations demonstrate similar levels of magnetization, developing magnetic fields no less than $22$ T during the 200 ps timescale of laser-target interaction. Figure~\ref{fig:bscan} summarizes the findings of all numerical convergence tests. The mass ratio is the parameter introducing the most uncertainty in the magnetic field amplitude. Still, in all cases the magnetic fields are strong enough to produce conditions with $\omega_{\rm ce} \tau_e >1$. {Particle depletion, while naturally appearing in the near-vacuum region in PIC approach, does not affect the main results - plasma profiles still converge decently, and the regions of Weibel magnetization ($n_e/n_{\rm cr}\sim 0.1$, $T_e\sim 1$ keV) are always resolved with at least $10^4$ electrons per Weibel mode wavelength.}

Finally, to estimate the growth of noise-level magnetic fields in the absence of the Weibel instability, we conduct a test simulation with a small 2D box ($7\mu \rm m\ \times 7 \mu \rm m$), no laser heating, and a uniform plasma with $n_e/n_{\rm cr}=0.1$, $T_e=1.2~\rm keV$, using the primary resolution parameters. The initial level of magnetic field fluctuations in this simulation (rms) is 0.3 T and stays below 0.6 T over 200 ps. These fields are small compared with the $50~\rm T$ magnetic fields in the ablation simulations and only act as seed fields.}

\begin{figure}
    \centering
    \includegraphics[width=\linewidth]{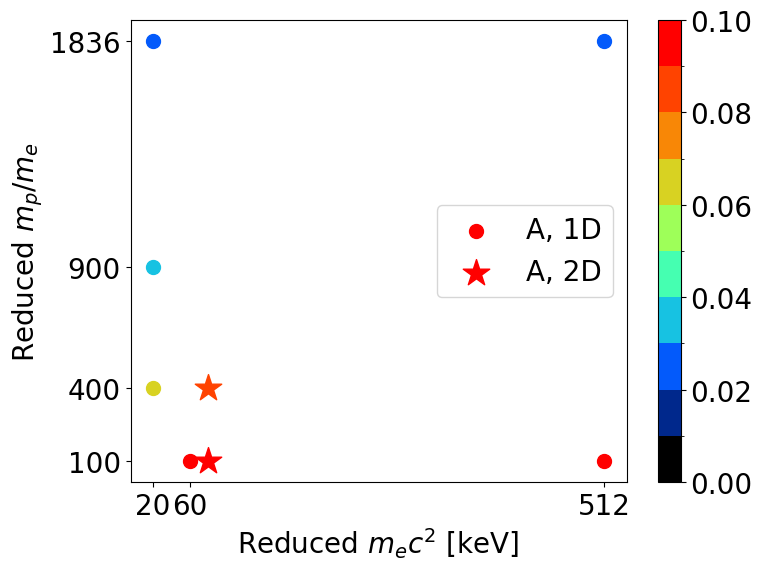}
    \caption{\normalsize{Electron temperature anisotropy levels for 1D and 2D simulations for various reduced $m_p/m_{e\ast}$ and $c_\ast$ parameters. In all considered cases, $A\geq 0.02$.}}
    \label{fig:anisotropiesA}
\end{figure}

\section{Anisotropy convergence tests}\label{sec:convergence}

It is important to note that we utilize reduced mass ratio and speed of light parameters to make our simulations feasible, as is commonly done for the PIC simulations of high-energy-density plasma~\cite{Schoeffler2014,Schoeffler2016,Zhao2024,Fox2018}. One drawback is that the electron mean free path, and thus the anisotropy levels, are overestimated. We perform convergence studies with varied mass ratios ($m_p/m_{e\ast}=100,\,400,\,900, 1836$) and electron rest energies ($m_{e\ast}c_\ast^2 = 20,\,60,\,512$ keV), which show roughly consistent results for the fastest-growing mode. Anisotropy levels decrease as the mass ratio and speed of light approach realistic values, but they do not vanish completely. As shown in Fig.~\ref{fig:anisotropiesA}, the anisotropy remains at or above $A=0.02$ even in a full mass ratio and speed of light simulation. Therefore, we conclude that our qualitative results remain valid despite the use of reduced simulation parameters.

\section{Test of anisotropy development by laser module}\label{sec:aniso_test}

\begin{figure}
    \centering
    \includegraphics[width=1\linewidth]{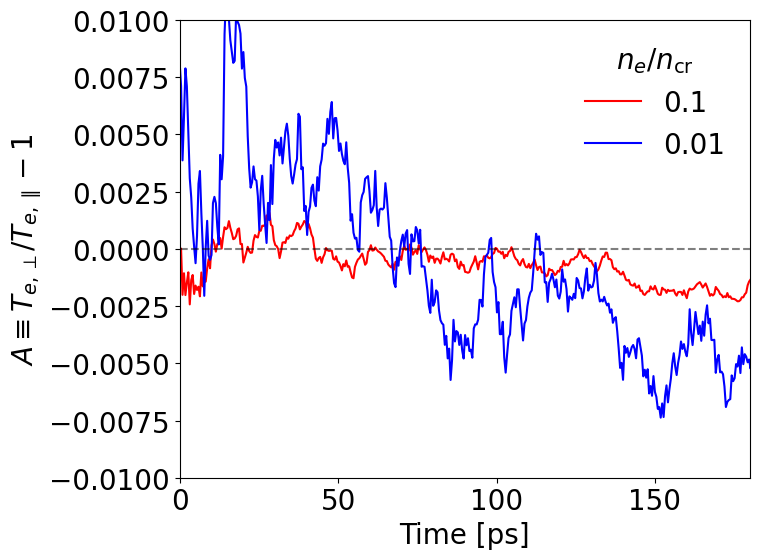}
    \caption{\normalsize{Anisotropy generated by laser heating for plasma parameters relevant to our 1D/2D PSC simulations for $n_e=0.1 n_{\rm cr}$ (red) and $n_e=0.01 n_{\rm cr}$ (blue). The anisotropy generated directly from the laser heating is sign-alternating and negligible compared to the values observed in the primary ablation simulations ($A\approx 0.1$).}}
    \label{fig:anisotropiesB}
\end{figure}

\begin{figure*}
    \centering
    \includegraphics[width=1\linewidth]{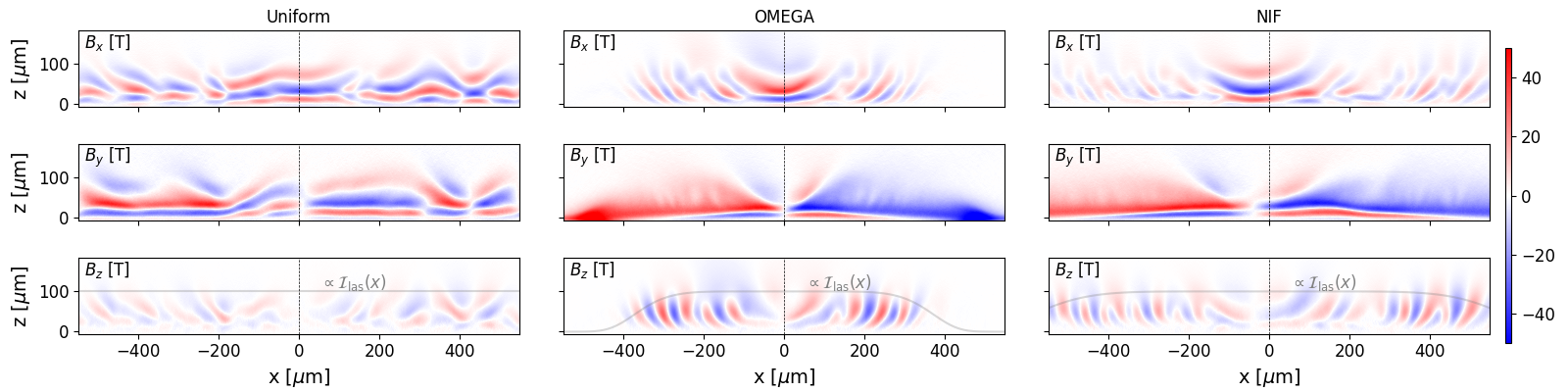}
    \caption{\normalsize{{The effect of the transverse laser intensity profile on magnetic field generation. Magnetic field components ($B_x$, $B_y$, $B_z$) are shown on the top, middle, and bottom rows for three laser intensity profiles: uniform beam (left), OMEGA beam (middle), and NIF beam (right). Gray lines on the bottom row depict normalized intensity profiles. The laser axis is highlighted by vertical black dashed lines.}}}
    \label{fig:laserprof1}
\end{figure*}

Another important validity check for our simulations is whether the laser heating module generates any plasma anisotropy on its own. While such a process may occur for a more general inverse Bremsstrahlung heating operator~\cite{Bendib1998}, our implementation uses a laser heating operator that only provides random kicks and therefore should not generate any anisotropy by design~\cite{Hyder2024}. To confirm the absence of any sizable level of anisotropy generated by the laser deposition module, we conduct the following tests. We consider a small 1D box of 1 $d_i$ size, resolved by 50 grid cells, with $10^5$ electrons per cell at the critical density. We fill the simulation box with a uniform Al plasma of either $0.1 n_{\rm cr}$ or $0.01 n_{\rm cr}$ electron densities, with the initial temperature $T_{e0}=60 \rm \, eV$ and flow velocity of 500 km/s — typical parameters of the expanding plasma around the region where the magnetic field emerges (see Fig.~1a of the primary manuscript). Periodic boundary conditions are applied to avoid plasma-boundary interaction effects. A laser with an intensity of $10^{14} \rm W/cm^2$ and wavelength of $1.064 \, \rm \mu m$ is used to heat the initially cold plasma. To isolate the effect of laser heating on plasma anisotropy, we turn off the electromagnetic field solver and collisions; therefore, the particles in the computational domain move ballistically until they are kicked by the laser heating module. It should be noted that simulations using the full field solver and binary collisions lead to similar conclusions. The simulation is conducted for 180 ps, the typical timescale of our primary simulations in the paper. Figure~\ref{fig:anisotropiesB} depicts the evolution of the mean electron temperature anisotropy inside the simulation box for the initial target densities of $0.1 n_{\rm cr}$ (red line) or $0.01 n_{\rm cr}$ (blue line). First, the maximum anisotropy level is $0.002$ for the $0.1 n_{\rm cr}$ case and $0.01$ for the $0.01 n_{\rm cr}$ case, as compared to $A \gtrsim 0.1$ in the manuscript. Next, the sign of the anisotropy changes over the course of the simulation, in contrast to the primary simulations in the manuscript. To determine whether the evolution of anisotropy can be characterized as a random walk, we apply the statistical Augmented Dickey–Fuller (ADF) test~\cite{ADF} to the anisotropy time series. The first 10 picoseconds are excluded since rapid laser heating occurs. We conclude that the laser heating module does not generate the anisotropy by itself. The resulting p-values for the presented anisotropy curves are 0.207 and 0.368 for $n_e=0.1\,n_{\rm cr}$ and $0.01\,n_{\rm cr}$, respectively. At the 95\% significance level, we therefore fail to reject the null hypothesis, meaning that the anisotropy curves are consistent with the random walk hypothesis.

\section{The role of laser heating}\label{sec:heating_role}

Despite the fact that the laser does not directly produce anisotropy, as shown above, the laser heating plays an important role in Weibel magnetogenesis. We conducted 1D simulations showing that turning off laser heating 0.1 ns into the simulation (rather than leaving it on) results in a very fast plasma cooldown driven by plasma expansion, and no Weibel instability develops. Thus, one of the roles of the laser heating is sustaining and, possibly, increasing the temperature of the expanding plasmas. {We also initialized our simulation from an analogous FLASH simulation at different output timings (0.02 and 0.1 ns), and we observed the formation of qualitatively similar magnetic filament structures.} Lastly, we conducted a 2D PSC simulation with no laser heating, initiated from an analogous FLASH simulation snapshot taken at 0.3 ns. Such plasma was hot enough and expanded fast enough to develop expansion-driven Weibel filaments of similar geometry, although the ultimate field amplitudes were smaller since the plasma was rapidly cooling down. We therefore conclude that the main role of laser heating in our simulations is to sustain high plasma temperature, mitigating collisional isotropization and driving a hot plasma expansion, which is the ultimate driver of the Weibel instability.

\begin{figure}
    \centering
    \includegraphics[width=1\linewidth]{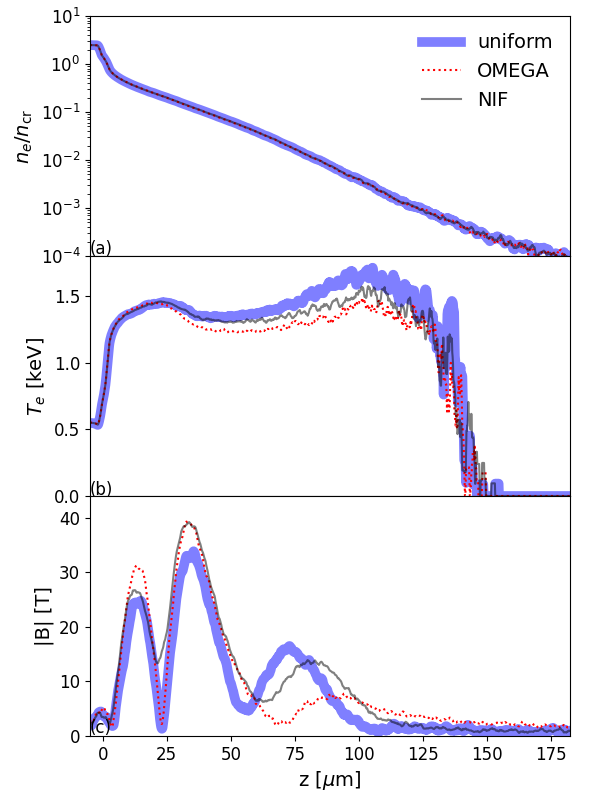}
    \caption{\normalsize{{Plasma conditions on the laser axis in simulations with three different transverse laser beam profiles (uniform, OMEGA, NIF). Magnetic field profiles are nearly identical despite a more complex overall magnetization picture in the OMEGA and NIF cases.}}}
    \label{fig:laserprof2}
\end{figure}

{
\section{The role of transverse laser intensity profile}\label{sec:laserspot}
In the main body of the manuscript, we consider either 1D simulations or 2D simulations with a uniform transverse laser profile. Realistic laser beams possess some structure in the transverse laser intensity distribution, which may influence magnetization physics. A detailed investigation of the effects of the transverse laser pulse profile deserves a separate study and is beyond the scope of this work. Below, we will justify the validity of using uniform laser intensity profiles and outline the main effects of the transverse laser intensity profile.

We conducted three auxiliary simulations: 2D simulations with a transverse box size of 1.5 mm and (1) a uniform laser beam profile, $\mathcal{I}_{\rm uniform} = \mathcal{I}_0=10^{14}~\rm W/cm^2$, (2) an OMEGA-like laser beam profile, $\mathcal{I}_{\rm OMEGA} = \mathcal{I}_0 \exp{\left[-\left(|x|/380{\mu \rm m} \right)^7 \right]}$, and (3) a NIF-like laser beam profile, $\mathcal{I}_{\rm NIF} = \mathcal{I}_0 \exp{\left[-\left(|x|/600{\mu \rm m} \right)^7 \right]}$. Figure~\ref{fig:laserprof1} depicts the resulting magnetic field structures at $t=150$ ps. The uniform case is nearly identical to the primary simulation discussed in the body of the manuscript. For finite laser spot sizes, the picture becomes more complicated. First, Biermann-like magnetic fields appear around $x \approx \pm 400 ~\mu \rm m$ early on in these simulations, even before the Weibel fields appear. Later in time, filamentary magnetic structures appear in both the OMEGA and NIF cases, with oblique magnetic structures seen around the edges of the pulse. Eventually, strong magnetic fields similar to the ones observed in the uniform intensity case appear near the laser axis (see the $B_x$ component for the OMEGA and NIF cases, near $x=0$ and $z=50 ~\mu \rm m$). We conclude that the general 2D case featuring an isolated laser spot with a nonuniform intensity profile also produces Biermann fields, leading to a more complex picture. Still, if one compares plasma profiles and magnetic fields near the laser axis ($x=0$) among the three different laser pulse profiles, a nearly identical picture emerges. Figure~\ref{fig:laserprof2} compares electron density, temperature, and magnetic field magnitude on the laser axis at t=150 ps. Despite the complex magnetogenesis in the non-uniform intensity profile, on-axis magnetization is nearly identical to the uniform transverse laser profile. Therefore, we conclude that the expansion-driven Weibel process plays an important role in magnetogenesis for realistic laser spot profiles. For direct-drive inertial fusion experiments aiming at on-target laser intensity variations of no more than $1\%$~\cite{craxton2015}, we predict the expansion-driven Weibel to be the dominant process, as our auxiliary simulations show, with the magnetic field profiles nearly identical to the left column of Fig.~\ref{fig:laserprof1}.}

{
\section{Analytical model of temperature anisotropy development in general sheared and compressive flows}\label{sec:analytics}

The paper’s main result, the emergence of flow-driven electron temperature anisotropy and subsequent Weibel magnetization in high-energy-density plasmas, can also be derived analytically. Below we derive the anisotropy growth rate for representative flows (planar, isotropic Hubble-like, and radial expansion from a spherical surface), introduce a simple dynamical model that includes collisional isotropization, and estimate the peak anisotropy, the number of collisionless electron Weibel e-foldings per collisional isotropization time and per the characteristic simulation times considered in the paper.

Let us start from the Lagrangian form of the evolution equation for the electron pressure tensor, obtained from the Vlasov equation under the assumption of zero electromagnetic forces:

\begin{equation}
    \frac{D}{D t}P_{ij}+ P_{ij} \nabla \cdot \mathbf{u} + P_{ik} \frac{\partial u_j}{\partial x_k}+ P_{jk} \frac{\partial u_i}{\partial x_k}+\frac{\partial Q_{ijk}}{\partial x_k} = 0.
\end{equation}

\noindent Summation over three dimensions is assumed for the terms with repeating indices (Einstein's notation). Here, $P_{ij} \equiv m_e \int d^3 v f_e(\mathbf{v})(v_i-u_i)(v_j-u_j)$ is the electron pressure tensor, $f_e(\mathbf{v})$ is the electron distribution function, $\mathbf{u} \equiv n^{-1} \int d^3 v f_e(\mathbf{v})\mathbf{v}$ is the mean flow speed, $n \equiv \int d^3 v f_e(\mathbf{v})$ is the number density of electrons, and $Q_{ijk}\equiv m_e \int d^3 v f_e(\mathbf{v})(v_i-u_i)(v_j-u_j)(v_k-u_k)$ is the heat flux tensor. In what follows, we ignore heat flux effects ($Q_{ijk}=0$) and define the electron temperature tensor as $T_{ij} \equiv P_{ij}/n$. Using the continuity equation, we obtain the following expression for the electron temperature tensor:

\begin{equation}
    \frac{D}{D t}T_{ij}+  T_{ik} \frac{\partial u_j}{\partial x_k}+ T_{jk} \frac{\partial u_i}{\partial x_k} = 0.
\end{equation}

\noindent This equation governs the evolution of the temperature tensor within the fluid element as a result of shear flows ($\partial u_i/\partial x_j$). It demonstrates that under collisionless conditions, a wide variety of plasma flows will drive temperature anisotropy. In fact, it is a very restricted class that \textit{does not} drive anisotropy.

For a diagonal electron temperature tensor, we find:

\begin{equation}
    \frac{D}{Dt} T_{\alpha \alpha}  + 2 T_{\alpha \alpha} \frac{\partial u_\alpha}{\partial x_\alpha} = 0,
\end{equation}

\noindent where we now indicate components $\alpha \in {x,y,z}$ for Cartesian or $\alpha \in {r,\theta,\phi}$ for spherical geometry, with the repetition not indicating summation. This result shows how the separate components of $T_{\alpha\alpha}$ evolve in the absence of collisions, and more importantly, how $\partial u_\alpha /\partial x_\alpha$ terms drive anisotropy. The temperature anisotropy is driven whenever the very restrictive condition $\partial u_x / \partial x = \partial u_y / \partial y = \partial u_z / \partial z$ ($\partial u_r / \partial x_r = \partial u_\theta / \partial x_\theta = \partial u_\phi / \partial x_\phi$) is violated.

We can now consider various geometries for concreteness. The simplest case is a one-dimensional planar flow ($u_z = a z$, $u_x=u_y=0$, $a=\mathrm{const}>0$), which is the primary case considered in the manuscript. Computing the velocity gradient for this flow in Cartesian coordinates ($\partial u_i/\partial x_j = a\delta_{iz}\delta_{jz}$, where $\delta_{ij}$ is the Kronecker delta) yields the following evolution equations for the diagonal elements:

\begin{eqnarray}
    \frac{D}{D t}T_{xx} = \frac{D}{D t}T_{yy}=0, \label{eq:planar1}\\
    \frac{D}{D t}T_{zz} + 2a T_{zz} =0. \label{eq:planar2}
\end{eqnarray}

\noindent It is evident that planar flow generates anisotropy. Defining $A\equiv 1 - T_{zz}/T_{\perp}$; $T_{\perp} \equiv (T_{xx}+T_{yy})/2$, the generation rate may be written as:

\begin{equation}
    \frac{D}{D t}A =  2a (1-A).
    \label{eq:dAdt}
\end{equation}

\noindent Now, if one considers a radial Hubble-like flow ($u_r = a r$, $u_{\theta}=u_{\phi}=0$, $a>0$), then, after computing the velocity gradient in spherical coordinates ($\partial u_i/\partial x_j = a\delta_{ij}$), one finds that there is no anisotropy generation from an initially isotropic distribution, since:

\begin{eqnarray}
    \frac{D}{D t}T_{rr} + 2aT_{rr} = 0, \label{eq:3Dhubble1} \\ 
    \frac{D}{D t}T_{\theta \theta} + 2aT_{\theta \theta} = 0, \label{eq:3Dhubble2}\\
    \frac{D}{D t}T_{\phi \phi} +  2a T_{\phi \phi} =0. \label{eq:3Dhubble3}
\end{eqnarray}

\noindent However, if one considers a more realistic scenario, such as a radial ablation flow from spherical target surface, and models it, e.g., as $u_{r} = a (r-R)$ for $r>R$ and $u_{r} =0$ for $r\leq R$, with $u_{\theta}=u_{\phi}=0$, $a>0$ and $R>0$ ($R$ is the target radius), one obtains a more complex velocity gradient ($\partial u_r/\partial x_r = a$, $\partial u_\theta/\partial x_\theta =\partial u_\phi/\partial x_\phi = a(r-R)/r$, and $\partial u_i/\partial x_j =0$ for $i\neq j$), resulting in the following temperature evolution equations:

\begin{eqnarray}
    \frac{D}{D t}T_{rr} + 2aT_{rr} = 0, \label{eq:3Dablation1} \\ 
    \frac{D}{D t}T_{\theta \theta} + 2a\frac{r-R}{r}T_{\theta \theta} = 0, \label{eq:3Dablation2} \\
    \frac{D}{D t}T_{\phi \phi} +  2a\frac{r-R}{r} T_{\phi \phi} =0. \label{eq:3Dablation3}    
\end{eqnarray}

\noindent We note that these equations recover the planar and isotropic Hubble-like flow limits for $|r-R|/R\ll 1$ and $r\gg R$, respectively. Thus, the ablation flow is effectively isotropic (no anisotropy generation) far from the target, but behaves as a planar flow (drives anisotropy) close to the target surface. In our simulations, Weibel growth occurs within $\sim100~\mu\mathrm{m}$ of the target, while typical direct-drive inertial fusion targets have radius $R\geq1~\mathrm{mm}$~\cite{craxton2015}. Therefore, the temperature tensor evolves according to the planar flow equations (Eqs.~\ref{eq:planar1},\ref{eq:planar2}), taking $r\approx R$ in Eqs.~\ref{eq:3Dablation2},\ref{eq:3Dablation3}. The anisotropy generation rate follows from Eq.~\ref{eq:dAdt}.

We conclude that temperature anisotropy can arise in a broad range of flows, including flows that are isotropic in angle (e.g., $\partial/\partial\theta=\partial/\partial\phi=0$), whenever the velocity gradient tensor has a nonisotropic component ($\partial u_i/\partial x_j - a\delta_{ij}\neq 0$). Radial ablation flow from a spherical target surface is a representative example.

\begin{figure}
    \centering
    \includegraphics[width=\linewidth]{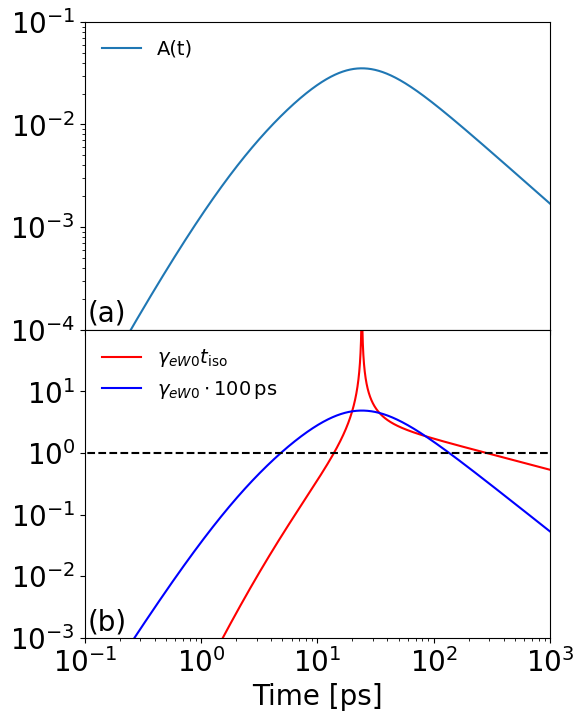}
    \caption{(a) Evolution of the electron temperature anisotropy $A$ for typical HED plasma parameters simulated in the manuscript (Al$^{13+}$, $n_e=10^{20}~\rm cm^{-3}$, $T_e=1~\rm keV$, $L_0 = 5~\rm \mu m$). (b) Evolution of the number of collisionless Weibel e-foldings per isotropization time and per typical simulation time.}
    \label{fig:analytics}
\end{figure}

In a collisional plasma, temperature anisotropy is smeared out by particle collisions. Following Ref.~\cite{Schoeffler2020} (Eqs.~4–6 therein) in the small anisotropy limit ($A\ll 1$), the relaxation rate is

\begin{align}
\frac{dA}{dt} &=
-\frac{6}{5}\!\left(1+\frac{1}{\sqrt{2}\,Z}\right)\nu_{ei}\,A, \\
\nu_{ei} &= 2.91\times10^{-6}\, Z\,\ln\Lambda\, \frac{n_e\,[\mathrm{cm^{-3}}]}{T_{e,\perp}^{3/2}[\mathrm{eV}]}\,[\mathrm{s^{-1}}].
\label{eq:colliso}
\end{align}

\noindent Here $\nu_{ei}$ is the electron–ion collision frequency evaluated using the perpendicular temperature~\cite{NRL}.

To predict the time evolution of the electron temperature anisotropy, we need to specify a representative flow for high-energy-density systems. One option is the self-similar analytical solution for a nonrelativistic, quasi-neutral, collisionless plasma evolving purely under electrostatic forces \cite{dorozhkina1998exact,bellei2012electron}. This model describes the adiabatic evolution of a thermal plasma, which is a reasonable approximation for weakly radiating HED plasmas after the laser pulse ends~\cite{schaeffer2016characterization}. For a one-dimensional expansion of an initially stationary Gaussian plasma plume with scale length $L_0$ and a Maxwellian velocity distribution characterized by the sound speed $C_S$, the expansion speed can be written as (e.g., \cite{dorozhkina1998exact}, Eqs.~11–13):

\begin{equation}
    u_z = z \cdot \frac{C_S^2 t}{C_S^2t^2+L_0^2}.
\end{equation}

\noindent The planar flow assumptions apply here. Substituting $\partial u_z/\partial z$ into Eq.~\ref{eq:dAdt} and combining with Eq.~\ref{eq:colliso}, we obtain:

\begin{equation}
    \frac{dA}{dt} = \frac{2C_S^2 t}{C_S^2t^2+L_0^2}(1-A)-\frac{6}{5}\left(1+\frac{1}{\sqrt{2}Z}\right)\nu_{ei}A.
    \label{eq:masteraniso}
\end{equation}

\noindent Solving this equation numerically, we obtain $A(t)$ and two quantities relevant for Weibel magnetization: the number of collisionless electron Weibel e-foldings per isotropization time $t_{\rm iso}\equiv |A/(dA/dt)|$, $\gamma_{eW0}t_{\rm iso}$, and the number of e-foldings over a typical simulation time (100 ps), $\gamma_{eW0}\cdot 100~\rm ps$. Figure~\ref{fig:analytics} shows the results for the parameters of a typical HED plasma simulated in the manuscript (Al$^{13+}$, $n_e=10^{20}~\rm cm^{-3}$, $T_e=1~\rm keV$, $L_0 = 5~\rm \mu m$). The peak anisotropy $A\approx 0.04$ is reached within $\approx 30$ ps, while $\gamma_{eW0}t_{\rm iso}$ and $\gamma_{eW0}\cdot 100~\rm ps$ remain well above one for $\approx 200$ ps. Thus, multiple Weibel e-foldings are expected on sub-nanosecond timescales, consistent with the simulations.
}

\end{document}